\documentclass[preprint,12pt,5p,twocolumn]{elsarticle}

\usepackage{longtable}
\usepackage{amssymb}
\usepackage{amsthm}
\usepackage{dcolumn}
\usepackage{graphicx}

\journal{Journal of Magnetism and Magnetic Materials}

\begin{document}

\begin{frontmatter}

\title{Antiferromagnetic Heisenberg S=5/2 spin chain compound SrMn$_2$V$_2$O$_8$}

\author{Sandra K.~Niesen}
\author{Oliver Heyer}
\author{Thomas Lorenz}
\author{Martin Valldor}

\address{Physikalisches Institut, Universit\"{a}t zu
K\"{o}ln, Z\"{u}lpicher Str. 77, 50937 K\"{o}ln, Germany}

\begin{abstract}
Large single crystals of the new compound SrMn$_2$V$_2$O$_8$ have been grown by the floating-zone method. This transition-metal based oxide is isostructural to SrNi$_2$V$_2$O$_8$, described by the tetragonal space group $I$4$_1cd$. Magnetic properties were investigated by means of susceptibility, magnetization, and specific heat measurements.
The title compound behaves like a one-dimensional magnetic system above the ordering temperature ($T_N$ = 43~K). The magnetic ground state can be described as a classical long-range ordered antiferromagnet with weak anisotropy. 
\end{abstract}

\begin{keyword}
 metal oxide \sep low dimensional \sep antiferromagnet 

\end{keyword}
\end{frontmatter}

\section{Introduction}
\label{sec:Intro}
Low-dimensional magnetic systems are commonly studied due to their interesting magnetic properties. For small spin values ($S=1/2$~or~1), 
the groundstate and the low-lying excitations are often dominated by strong quantum fluctuations, while a more classical behavior is expected 
for systems with larger spins. Up to now only few low-dimensional magnets containing Mn$^{2+}$ (d$^5$) with high spin $S=5/2$ have been studied, \textit{e.g.}\ clinopyroxene CaMnGe$_2$O$_6$ \cite{doi:10.1016/j.jssc.2008.08.014} or the spin-ladder compound BaMn$_2$O$_3$ \cite{PhysRevB.83.024418}.
In this context, compounds with the general formula $AM_2X_2$O$_8$ ($A$~= Ba, Sr, Pb; $M$ = Cu, Co, Ni, Mn; $X$ = V, As) are of particular interest. 
Depending on the transition-metal ion, different spins are realized and the structure contains screw chains of octahedrally coordinated $M^{2+}$ ions along the $c$ axis of the tetragonal structure. These chains are spatially separated by a nonmagnetic matrix, resulting in a quasi-1D magnetic system. The magnetic properties can be tailored by the used transition metal ions, \textit{e.g.} BaCu$_2$V$_2$O$_8$ ($S = 1/2$) is a 1D large spin gap system \cite{PhysRevB.69.220407}, PbNi$_2$V$_2$O$_8$ is an $S=1$ Haldane spin gap system \cite{PhysRevB.77.100401}, and BaCo$_2$V$_2$O$_8$ \cite{wich86, He2006}  as well as SrCo$_2$V$_2$O$_8$ \cite{muebu94, PhysRevB.73.212406} ($S = 3/2$) exhibit large magnetic anisotropy. 
The Heisenberg $S=5/2$ system BaMn$_2$V$_2$O$_8$ shows low-dimensional behavior around 170~K but finally orders antiferromagnetically at T$_N=37$~K \cite{muebu92, he07}. This N\'eel state is a result of the 3D coupling between the spin screw chains. Dzyaloshinskii-Moriya interactions \cite{dzyaloshinskii, PhysRev.120.91} were used to explain a spin canting \cite{he07}.
In the case of $M$~=~Co and $X$~=~V it is possible to have Sr or Ba at the $A$ site \cite{lejay}; in analogy, Sr might enter the $A$ site in BaMn$_2$V$_2$O$_8$ as well.
Thus, large single crystals of SrMn$_2$V$_2$O$_8$ were prepared and the crystal structure and the basic physical properties of this new compound are presented below.

\section{Experimental}
\begin{figure}
	\centering
		\includegraphics[width=0.8\linewidth]{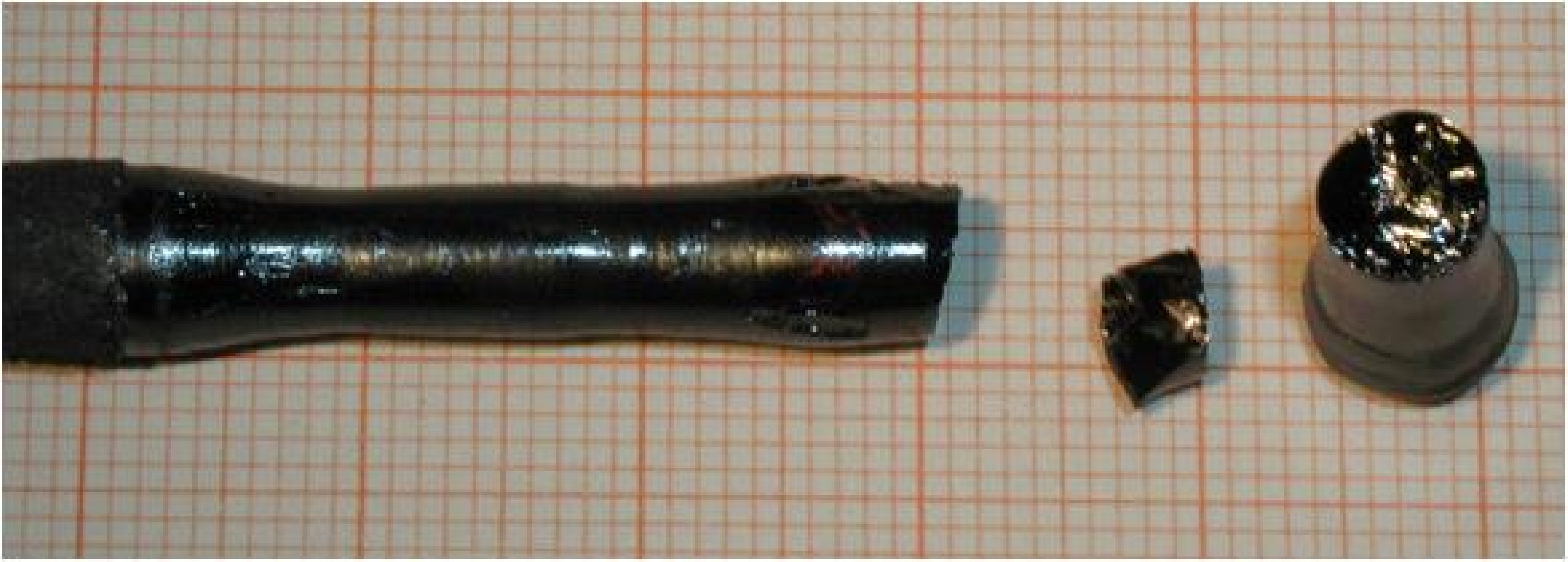}
		\caption{The as-grown single crystal of BaMn$_2$V$_2$O$_8$. A cleaved surface is shiny and smooth.}
	\label{fig:photo}
\end{figure}
As a precursor for the main synthesis, SrV$_2$O$_6$ was synthesized by a solid state reaction using a mixture of SrCO$_3$ (99.99 \% Alfa Aesar) and V$_2$O$_5$ (99.5~\% Strem Chemicals) in a 1:1 molar ratio. A Pt-crucible served as reaction vessel and the temperature was kept at 1000\symbol{23}C overnight. Subsequently, SrV$_2$O$_6$ was mixed with MnO (99 \% Aldrich) in the molar ratio of 1:2 and after homogenization in an agate motar the powder mixture was pressed into a bar and a seed.
A four mirror image furnace (FZ-T-10000-H-VI-VP, Crystal Systems Inc.) was used for the single crystal growth by means of the floating-zone technique. Ar, purified over hot metallic Ti, was chosen as reaction atmosphere.
For powder X-ray diffraction the Bragg-Brentano geometry was applied in a D5000 Stoe diffractometer with CuK$\alpha_{1,2}$ radiation ($\lambda_1=1.54056$~\AA, $\lambda_2=1.54439$~\AA) and a position sensitive detector.
The single crystal data acquisition was performed with a Bruker X8 APEX diffractometer at room temperature working with a MoK$\alpha_{1,2}$ ($\lambda_1$ = 0.70930~\AA, $\lambda_2$ = 0.71359~\AA) X-ray source. The obtained data were empirically absorption corrected using the softwares X-RED \cite{XRED} and X-SHAPE \cite{XSHAPE}. The subsequent refinement was completed with the JANA2000 \cite{jana} software package.

For magnetic measurements a SQUID MPMS-XL magnetometer from Quantum Design was used with magnetic fields up to 7~T in a temperature range 2-300~K.
Specific heat measurements were performed via thermal relaxation in a Quantum Design PPMS. The temperature range was 2-300~K and the maximum magnetic field reached 14~T.

\section{Results and Discussion}
\label{sec:Res}
\subsection{Crystal Structure}

\begin{figure}
	\centering
	\includegraphics[width=0.99\linewidth]{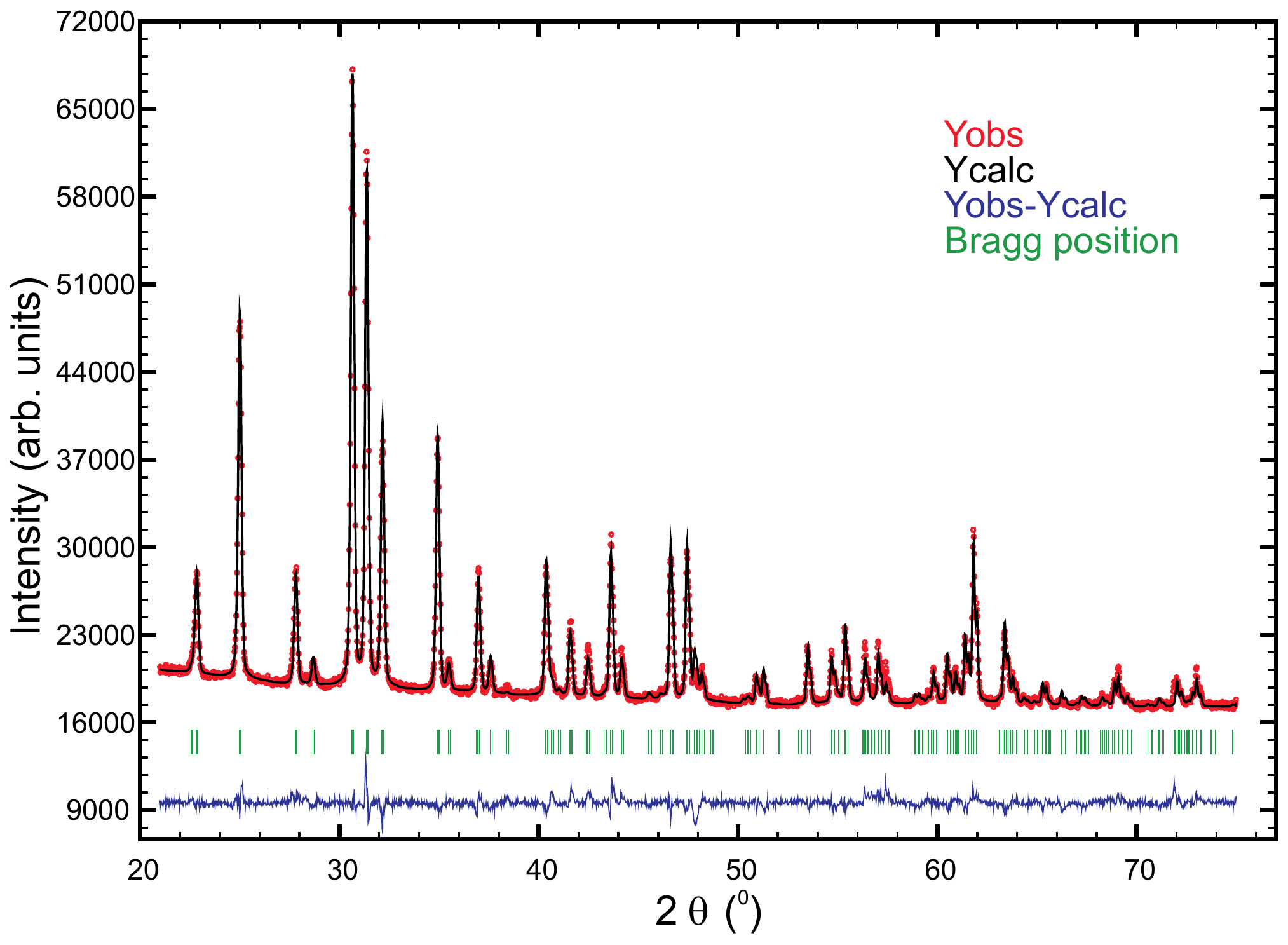}
	\caption{Powder X-ray Diffraction intensities (open circles) compared with the expected Bragg intensities (vertical lines) of a structure model (full line) based on the single crystal data.}
	\label{fig:XRPD}
\end{figure}

\begin{figure}
	\centering
		\includegraphics[width=0.99\linewidth]{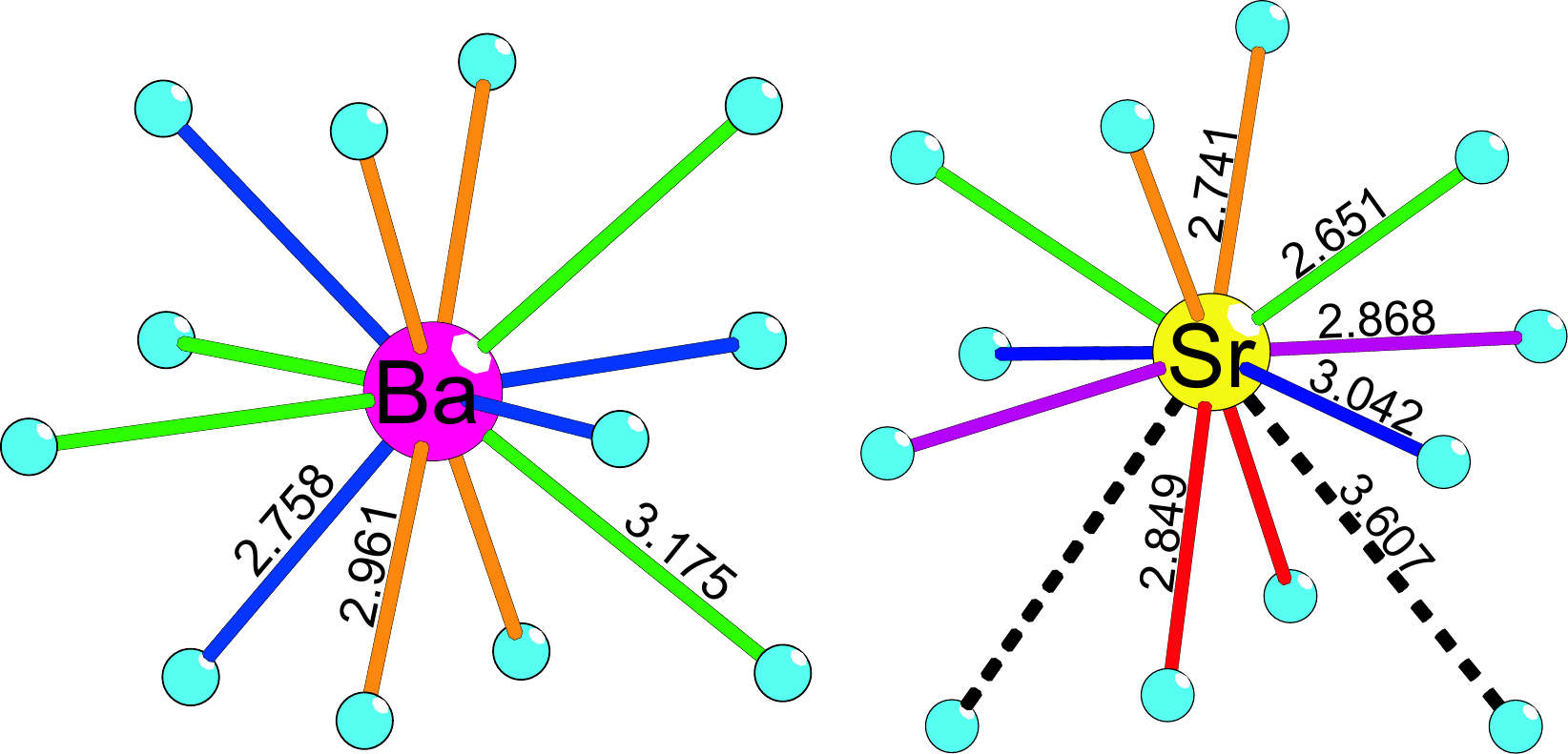}
	\caption{Comparison between the alkaline-earth cubeocahedra of BaMn$_2$V$_2$O$_8$ (left) \cite{muebu92} and SrMn$_2$V$_2$O$_8$ (right). For each polyhedron, metal-oxygen distances are labeled in \AA\ and equal distances have equal colors.}
	\label{fig:Polhyhedra}
\end{figure}

The resulting crystal from the mirror furnace is shown in Fig.~\ref{fig:photo}. 
A part of the cm-sized single crystal was crushed into a dark orange powder for powder x-ray diffraction.  
Using the tetragonal noncentrosymmetric space group \textit{I}4$_1$\textit{cd} (No. 110) it was possible to index all significant intensities in the powder X-ray data (Fig.~\ref{fig:XRPD}). The refined cell parameters are $a=12.4422(9)$~\AA~ and $c=8.6833(6)$~\AA, as obtained from a Rietveld refinement with Fullprof 2000 \cite{fullprof}. In these powder data no impurities were detected (Fig.~\ref{fig:XRPD}), \textit{i.e.}, the crystal is at least 95~\% pure.
The resulting parameters in the refined structural model from single crystal diffraction are presented in Tab.~\ref{tab:structure}.
The refinement was made on F$^2$ incorporating 59 parameters and 1494 measured intensities, of which 1324 were larger than 3$\sigma$. After reaching convergence, the maximum peak and hole in the difference Fourier map were 0.49 and -1.09 $e^-$/\AA$ ^3$, respectively.\begin{figure}
	\centering
		\includegraphics[width=0.99\linewidth]{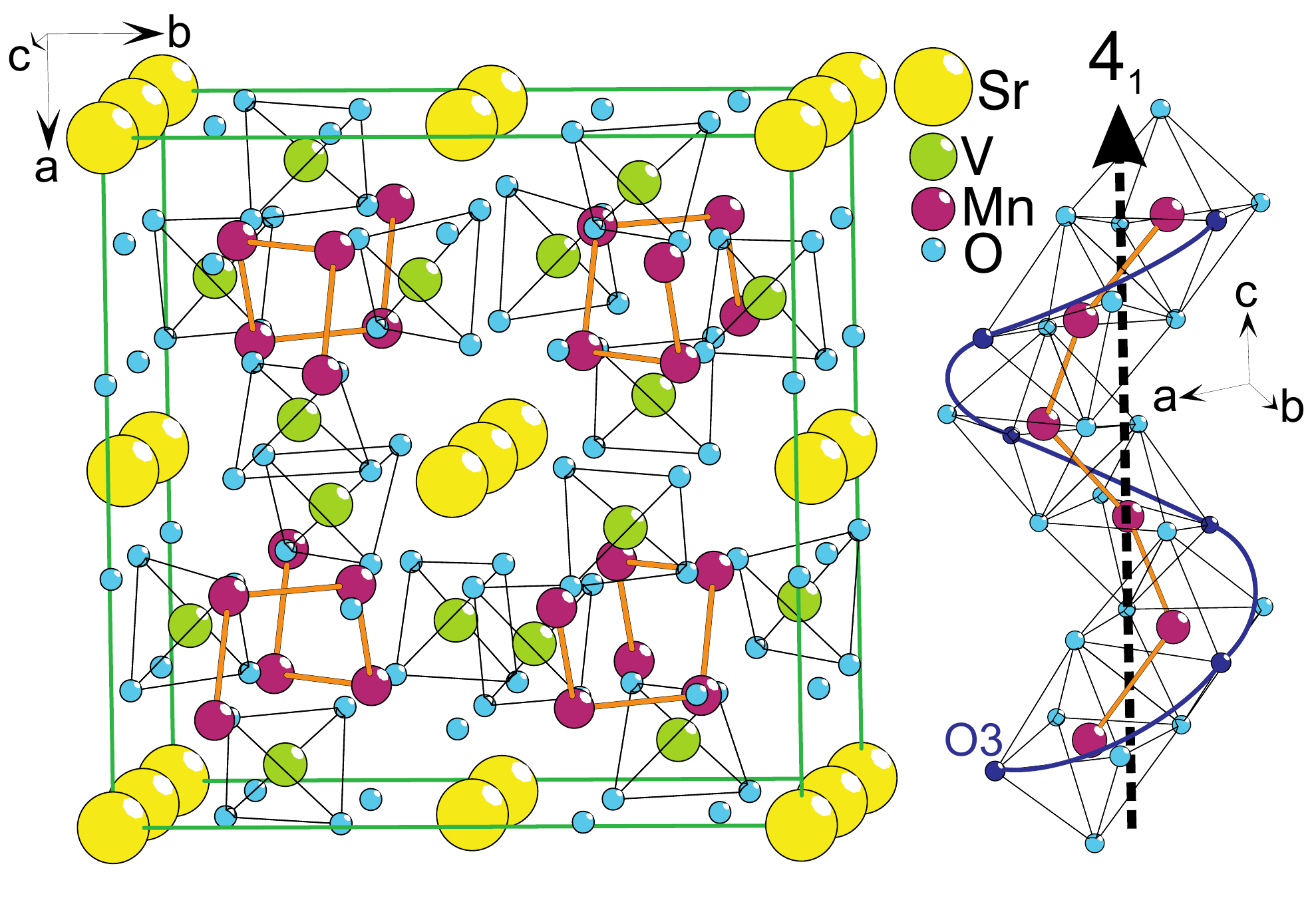}
	\caption{Perspective view of the unit cell (left) and the MnO spiral representing the 1D subsystem (right). O3 positions are connected by a full line to highlight the 4$_1$ screw axis, which itself is marked by the dashed arrow.}
	\label{fig:structure}
\end{figure}
SrMn$_2$V$_2$O$_8$, displayed in the left part of Fig.~\ref{fig:structure}, is isostructural to SrNi$_2$V$_2$O$_8$  and its detailed structure description can be found in \textit{e.g.} Ref.~\cite{muebuwich}. Here, we confine to a short presentation of the main structural features, in comparison to the isostructural compound BaMn$_2$V$_2$O$_8$ and with respect to the discussion of the magnetic properties.\\
The polyhedral building blocks of the structure are: a 12-fold oxygen coordinated Sr forming a distorted cubeoctahedron, a tetrahedrally coordinated V, and a 6-fold coordinated Mn displaced from the centre of an octahedron. The octahedra share edges and constitute quasi-1D chains along the $c$ axis, described by the 4$_1$ screw axis symmetry (Fig. \ref{fig:structure} right). The mean distance between the chains is $6.2211$~\AA~($=a/2$) (Fig.~\ref{fig:structure} left).
The alkaline-earth site ($A$) in SrMn$_2$V$_2$O$_8$ and BaMn$_2$V$_2$O$_8$ seems to play an important role considering the crystal symmetry. Comparing the $A$-O bond distances (Fig.~\ref{fig:Polhyhedra}) it is apparent, that Sr is too small for the site. The Sr-O bond lengths are in the range $2.647-3.601$~\AA \ and result in a bond valence sum (BVS) of $+1.602$ (Ba: +2.12). Sr is displaced in its cubeoctahedral environment almost reaching a coordination of ten, whereas the larger Ba is found almost in the center.
According to BVS calculation, the formal oxidation states of Mn are $+2.262$ for SrMn$_2$V$_2$O$_8$ and +2.32 for BaMn$_2$V$_2$O$_8$.
The notable deviations from the expected values, $+2$ for both ions, indicate a covalent contribution to the bonding character. In addition to the probable charge transfer between O and V, at least the covalence of Mn can participate in coloring these compounds.
Further consequences for SrMn$_2$V$_2$O$_8$ are: a breaking of the centro symmetry ($a$ glide plane) causing different values of the two nearest-neighbor Mn-O-Mn bond angles (Sr: 86.55$^\circ$; 86.12$^\circ$ ) instead of one (Ba: 86.93$^\circ$). 
In addition, the $a$ unit cell parameter and, hence, the mean interchain distance is significantly shorter in SrMn$_2$V$_2$O$_8$ (6.2309~\AA) than in  BaMn$_2$V$_2$O$_8$ (6.2781~\AA). The shorter interchain distance might result in a stronger interchain coupling.
 
\subsection{Magnetic Properties}
\begin{figure}
	\centering
		\includegraphics[width=0.99\linewidth]{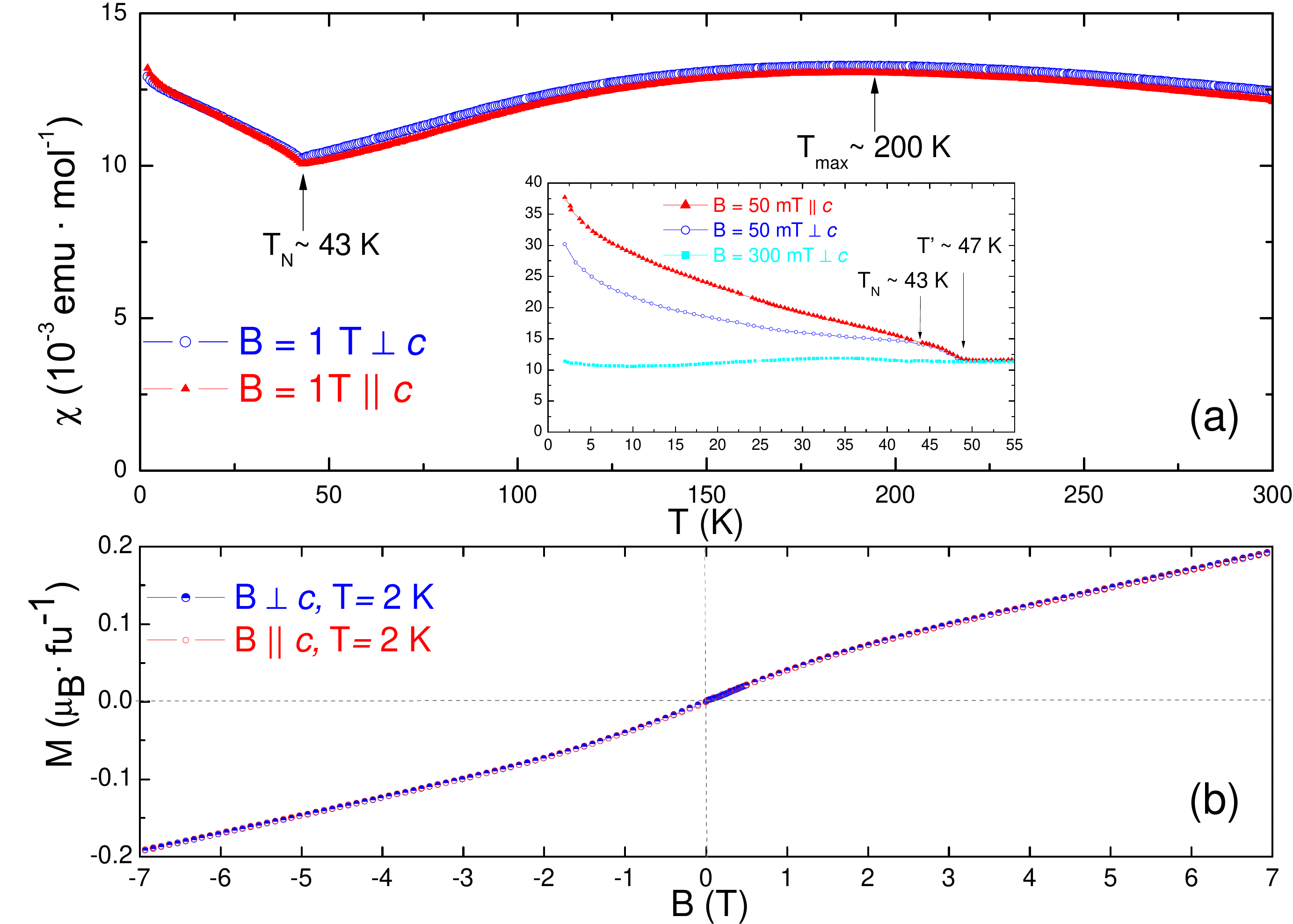}
	\caption{(a) Magnetic susceptibility $\chi$ of SrMn$_2$V$_2$O$_8$ for a magnetic field of 1~T applied either parallel or perpendicular to the $c$ axis. The N\'eel temperature is marked by T$_N$. The inset displays $\chi(T)$ measured in 300 and 50~mT showing an additional anomaly at 47~K. (b) Magnetization of SrMn$_2$V$_2$O$_8$ as a function of magnetic field applied parallel or perpendicular to $c$.}
	\label{fig:MChipaper}
\end{figure}
Magnetic susceptibility $\chi$ as a function of temperature and magnetization $M$ as a function of magnetic field are shown in Fig.~\ref{fig:MChipaper}. Both measurements have been done for field directions parallel and perpendicular to the crystallographic $c$ axis, which is the direction of the Mn-O chains. However, $\chi$ and $M$ show a nearly isotropic behavior. As the Mn$^{2+}$ ion with $3d^5$ configuration has a total angular momentum $L=0$ with a high spin of $S=5/2$, SrMn$_2$V$_2$O$_8$ is a Heisenberg system, where only weak anisotropies are expected.
A broad maximum in $\chi$ (Fig.~\ref{fig:MChipaper}a) centered around 200~K is observed for temperatures above T$_N$ (compared to 170~K in BaMn$_2$V$_2$O$_8$ \cite{he07}) confirming that the magnetic system is low dimensional, as expected from the crystal structure. 
Below T$_N$ the susceptibility shows a slight increase, which is typical for the transverse susceptibility $\chi_{\perp}$ of an antiferromagnet. 
The fact that the same $\chi_{\perp}$ and $M(B)$ are found for  both, a magnetic field applied perpendicular or parallel to the tetragonal $c$ axis of SrMn$_2$V$_2$O$_8$, means that a possible spin-flop field is extremely small or, in other words, the  Mn$^{2+}$ spins are practically isotropic.
In the magnetization (Fig.~\ref{fig:MChipaper}b) a linear increase is observed for higher fields, as expected for a canting of the moments in transverse fields well below the saturation field. In the maximum field of 7~T, only 0.2~$\mu_B/fu$ are observed, which correspond to only 2~\% of  the fully polarized magnetization of 10~$\mu_B/fu$. This agrees with a dominant antiferromagnetic coupling in SrMn$_2$V$_2$O$_8$. The lack of magnetic hysteresis negates ferro or ferri contributions to the ordered magnetic state in the main phase. 
The inset of Fig.~\ref{fig:MChipaper}a shows $\chi$ in low magnetic fields. Here, a second anomaly is observed close to 47~K, which is fully suppressed already at fields above 300~mT.
The disappearance of the 47~K anomaly suggests a possible ferromagnetic component, most likely originating from a magnetic impurity phase.
From a comparison of the magnetic moments in different fields and the S-shaped low-field magnetization curve (Fig.~\ref{fig:MChipaper}b) the amount of oriented moments and, hence, of the impurity phase can be roughly estimated to about 1~\%.
Two natural impurity candidates are Mn$_3$O$_4$ (hausmannite) \cite{dwight, chardon} and MnO \cite{seo}. 
Mn$_3$O$_4$ orders ferrimagnetically at 42~K, which is too small to explain the 47~K anomaly.
MnO is an antiferromagnet with T$_N=125$~K, but shows weak ferromagnetism as a nanoparticle with transition temperatures around 20~K \cite{seo}. Therefore it is difficult to identify one of these candidates as the measured impurity.

\subsection{Specific Heat}
\begin{figure}
	\centering
	\includegraphics[width=1.0\linewidth]{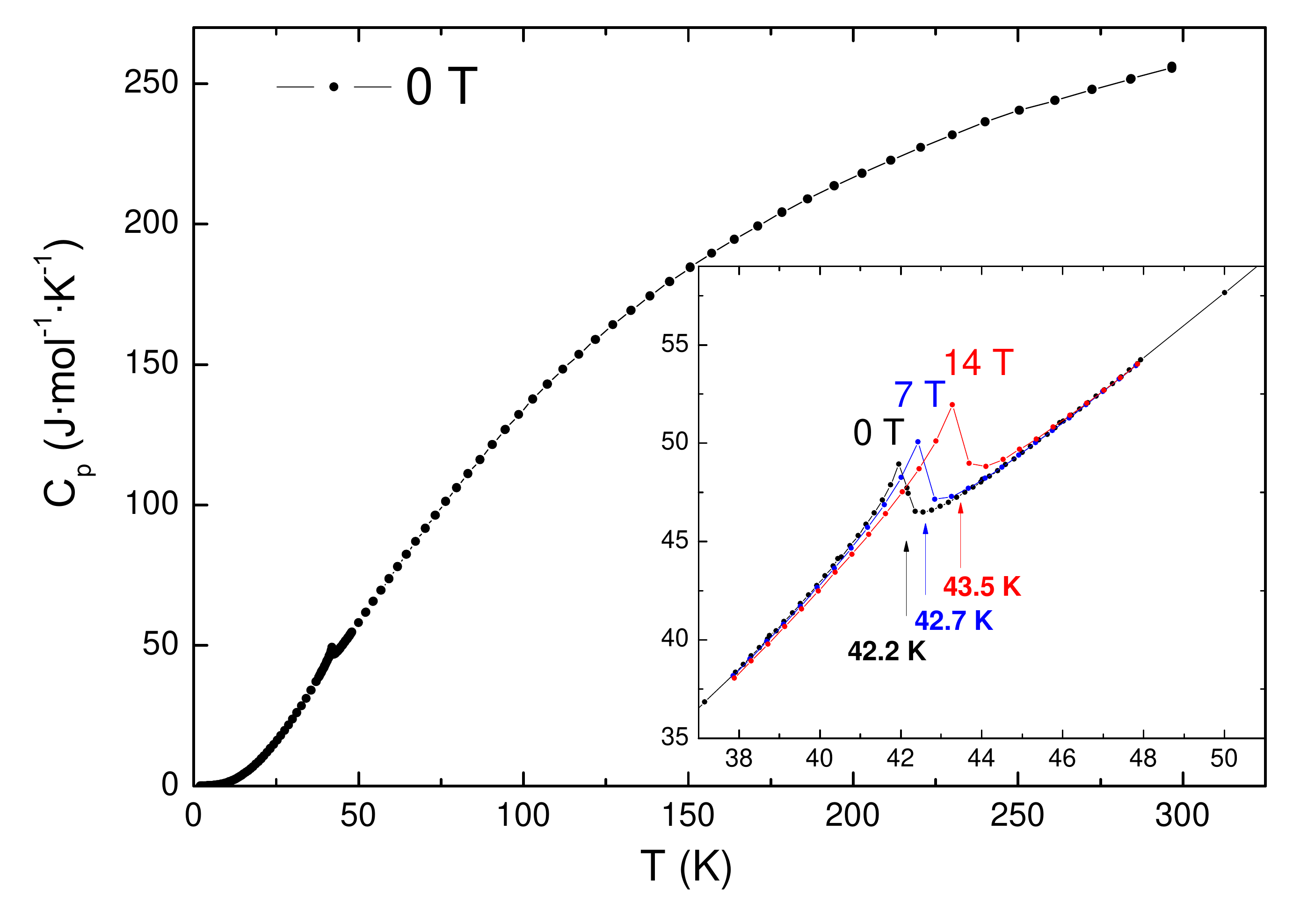}
	\caption{Specific heat as a function of temperature between 2 and 300~K. Inset: Magnification of the temperature range of the magnetic ordering in three different magnetic fields. The ordering temperatures are indicated by arrows.}
	\label{fig:Hc}
\end{figure}

In specific heat data only one anomaly is observed (Fig.~\ref{fig:Hc}). This coincides with the lower transition observed in the magnetic susceptibility around 43~K. The shape of this peak suggests a second-order phase transition, which is expected for a magnetic ordering. 
In an applied magnetic field the peak is slightly shifted to higher temperatures (inset of Fig.~\ref{fig:Hc}). The increase of $T_N$ amounts to about 70~mK/T and is in qualitative agreement with an estimation via the Ehrenfest relation $\frac{\partial T_N}{\partial H} = - T_N \frac{\Delta \frac{\partial M}{\partial T}}{\Delta c_p}$, which relates the field dependence of the N\'eel temperature to the respective anomalies observed in magnetization and specific heat at $T_N$. On decreasing temperature the slope $\partial M/\partial T$ changes from positive to negative at $T_N$, whereas the specific heat increases: thus, a positive field dependence of the transition temperature is expected. Note, that 
the 47~K anomaly observed in $\chi(T)$ for the lowest fields has no counterpart in the zero-field specific-heat data, what further confirms that this anomaly is no bulk feature of SrMn$_2$V$_2$O$_8$. 

\section{Summary}
\label{sec:Sum}
Large single crystals of the new compound SrMn$_2$V$_2$O$_8$ were successfully grown  by the floating-zone technique.
From crystal structure, susceptibility, magnetization, and heat capacity measurements it is clear that SrMn$_2$V$_2$O$_8$ belongs to the quasi-1D spin-5/2 antiferromagnets. The relatively high antiferromagnetic transition temperature around 43~K may be due to the relatively strong intra- and/or interchain magnetic coupling in the system.

\section*{Acknowledgement}
This work was supported by the Deutsche Forschungsgemeinschaft
through SFB 608.

\begin{table*}
  \caption{Single crystal data from X-ray diffraction. U$_{iso}$ and U$_{ij}$ are given in (10$^{-3}$ \AA$^2$) with standard deviations in parentheses.}
  \label{tbl:scd}
  \begin{tabular}{|c|c|c|c|}
    \hline
    \textbf{Atom, Wyck.} & \textbf{Coordinates} ($x,y,z$) & \textbf{Occupancy, U}$_{iso}$ & 	\textbf{U}$_{11}$, \textbf{U}$_{22}$, \textbf{U}$_{33}$, \textbf{U}$_{12}$, \textbf{U}$_{13}$, \textbf{U}$_{23}$\\
    \hline
    Sr, \textit{8a} & 0, 0, 0 & 1.0,  & 15.0(2), 13.6(2),\\
    & &14.8(1) & 15.6(2), 3.6(2),\\
    & & & 0, 0\\
		\hline
		V, \textit{16b} & 0.26406(4),   &  1.0,  & 7.4(2), 5.5(2),\\ 
		&0.07899(4),		&6.68(10)		&7.1(2), 0.2(2),\\
		&0.0779(1)		&		&-0.5(2), -0.2(2)\\
		\hline
		Mn, \textit{16b} & 0.33080(4),   & 1.0,  & 9.1(2), 8.8(2),\\  
		&0.33215(4),		&8.4(1)		&7.3(2), -1.0(2),\\
		&	0.21024(9)	&		&-0.6(2), 0.8(2)\\
		\hline
		O1, \textit{16b} & 0.1408(2),   & 1.0,  & 12(1), 8.5(9),\\ 
		&	0.4981(2),	&	12.5(6)	&17(1), -0.9(8), \\
		&	-0.0080(3)	&		&-1.9(9), -0.9(8)\\
		\hline
		O2, \textit{16b} & 0.1533(2),   & 1.0,  & 12(1), 16(1),\\  
		&0.6869(2),		&12.0(6)		&8.1(8), 4.2(9),\\
		&0.7009(3)		&		&0.7(8), -2.0(8)\\
		\hline
    O3, \textit{16b} & 0.3239(2),  & 1.0,  & 14(1), 10.2(9),\\  
    &	0.4995(2),	&	12.9(6)	&15(1), 1.8(8),\\
    &	 0.1710(3)	&		&-3.6(8), 2.2(8)\\
    \hline
    O4,\textit{16b} & 0.3328(2),   & 1.0,  & 14(1), 10(1),\\  
    &	0.1560(2),	&	11.1(6)	&9(1), -2.1(8),\\
    &	0.2124(3)	&		&-0.9(7), -0.7(7)\\
    \hline
    \multicolumn{4}{|l|}{Space group: \textit{I}4$_1$\textit{cd} (No.110), $a= 12.4422(9)$~\AA, $c= 8.6833(6)$~\AA, cellvolume$= 1350.697$~\AA$^3$,}\\
    \multicolumn{4}{|l|}{$Z=8$, $R$(obs)= 0.0268, $R_W$(obs)= 0.0666, $R$(all)= 0.0305, $R_W$(all)= 0.0677, }\\
    \multicolumn{4}{|l|}{ S(obs)= 1.20, S(all)= 1.15}\\
    \multicolumn{4}{|l|}{Unit cell parameters were adapted from the powder X-ray diffraction experiment.}\\
    \multicolumn{4}{|l|}{CSD No. 421422 \scriptsize(Data obtainable from FIZ, Karlsruhe, Abt. PROKA, 76344 Eggenstein-Leopoldshafen, Germany.)}\\
    \hline
  \end{tabular}
  \label{tab:structure}
\end{table*}

\begin{thebibliography}{10}

\bibitem{doi:10.1016/j.jssc.2008.08.014}
G.~Redhammer, G.~Roth, W.~Treutmann, W.~Paulus, G.~Andr$\rm\acute{e}$,
  C.~Pietzonka, G.~Amthauer, Magnetic ordering and spin structure in Ca-bearing clinopyroxenes
CaM$^{2+}$(Si, Ge)$_2$O$_6$, M = Fe, Ni, Co, Mn, Journ. of
  Solid State Chem. 181 (2008) 3163-3176.

\bibitem{PhysRevB.83.024418}
M.~Valldor, O.~Heyer, A.~C.~Komarek, A.~Senyshyn, M.~Braden, T.~Lorenz, Magnetostrictive N\'eel ordering of the spin-$\frac{5}{2}$ ladder compound BaMn$_{2}$O$_{3}$: Distortion-induced lifting of geometrical frustration,
  Phys. Rev. B 83~(2) (2011) 024418.

\bibitem{PhysRevB.69.220407}
Z.~He, T.~Ky\^omen, M.~Itoh, BaCu$_2$V$_2$O$_8$: Quasi-one-dimensional alternating chain compound with a large spin gap, Phys. Rev. B 69~(22) (2004) 220407.

\bibitem{PhysRevB.77.100401}
A.I.~Smirnov, V.N.~Glazkov, T.~Kashiwagi, S.~Kimura, M.~Hagiwara, K.~Kindo,
  A.Y.~Shapiro, L.N.~Demianets, Triplet spin resonance of the Haldane magnet  PbNi$_2$V$_2$O$_8$  with interchain coupling, Phys. Rev. B 77~(10) (2008) 100401.

\bibitem{wich86}
R.~Wichmann, H.~M\"uller-Buschbaum, Neue Verbindungen mit SrNi$_2$V$_2$O$_8$-Struktur: BaCo$_2$V$_2$O$_8$ und BaMg$_2$V$_2$O$_8$, Z.Anorg. Allg. Chem. 534 (1986) 153.

\bibitem{He2006}
Z.~He, T.~Taniyama, M.~Itoh, Large magnetic anisotropy in the quasi-one-dimensional system BaCo$_2$V$_2$O$_8$, Appl. Phys. Lett. 88 (2006) 132504.

\bibitem{muebu94}
D.~Osterloh, H.~M\"uller-Buschbaum, Zur Kenntnis von SrCo$_2$V$_2$O$_8$ und SrCo$_2$(AsO$_4$)$_2$, Z. Naturforsch. B 49 (1994) 923.

\bibitem{PhysRevB.73.212406}
Z.~He, T.~Taniyama, M.~Itoh, Antiferromagnetic-paramagnetic transitions in longitudinal and transverse magnetic fields in a SrCo$_2$V$_2$O$_8$ crystal, Phys. Rev. B 73 (2006) 212406.

\bibitem{muebu92}
M.~von Postel, H.~M\"uller-Buschbaum, Zur Kenntnis von Ba(MgZn)V$_2$O$_8$, BaMn$_2$V$_2$O$_8$ und Ba$_{1/2}$Sr$_{1/2}$Ni$_2$V$_2$O$_8$, Z.Anorg. Allg. Chem. 615 (1992) 97.

\bibitem{he07}
Z.~He, Y.~Ueda, M.~Itoh, Magnetic properies of the quasi-one-dimensional system BaMn$_2$V$_2$O$_8$, Solid State Commun. 141 (2007) 22.

\bibitem{dzyaloshinskii}
I.~Dzyaloshinskii, A thermodynamic theory of ''weak'' ferromagnetism of antiferromagnetics, J. Phys. Chem. Solids 4 (1958) 241.

\bibitem{PhysRev.120.91}
T.~Moriya, Anisotropic superexchange interaction and weak ferromagnetism, Phys. Rev. 120 (1960) 91.

\bibitem{lejay}
P.~Lejay, E.~Canevet, S.K.~Srivastava, B.~Grenier, M.~Klanjsek, C.~Berthier, Crystal growth and magnetic property of MCo$_2$V$_2$O$_8$ (M=Sr and Ba), Journ. of Crystal Growth (2011) doi:10.1016/j.jcrysgro.2011.01.016

\bibitem{XRED}
X-RED, STOE \& Cie GmbH, Darmstadt, Germany, v. 1.07 (1996).

\bibitem{XSHAPE}
X-SHAPE, STOE \& Cie GmbH, Darmstadt, Germany, v. 1.01 (1996).

\bibitem{jana}
http://www xray.fzu.cz/jana/jana.html.

\bibitem{fullprof}
J.~Rodr\'iguez-Carvajal, fullprof, Physica B 192 (1993) 55.

\bibitem{muebuwich}
R.~Wichmann, H.~M\"uller-Buschbaum, SrNi$_2$V$_2$O$_8$: ein neuer Strukturtyp der Erdalkali-Oxometallate, Revue de Chimie Minerale 23 (1986) 1.

\bibitem{dwight}
K.~Dwight, N.~Menyuk, Magnetic properties of Mn$_3$O$_4$ and the canted spin problem, Physical Review 119 (1960).

\bibitem{chardon}
B.~Chardon, F.~Vigneron, Mn$_3$O$_4$ comensurate and incommensurate magnetic structures, J. Magn. Magn. Mater. 58 (1986) 128.

\bibitem{seo}
W.~Seo, H.~Jo, K.~Lee, B.~Kim, S.~Oh, J.~Park, Size-dependent magnetic properties of colloidal Mn$_3$O$_4$ and MnO nanoparticles, Angew. Chem 116 (2004) 1135.

\end{thebibliography}
\end{document}